# Strong On-Chip Microwave Photon–Magnon Coupling Using Ultralow-Damping Epitaxial $Y_3Fe_5O_{12}$ Films at 2 Kelvin

Side Guo, Daniel Russell, Joseph Lanier, Haotian Da, P. Chris Hammel, and Fengyuan Yang

Department of Physics, The Ohio State University, Columbus, OH 43210

Abstract

$Y_3Fe_5O_{12}$ is arguably the best magnetic material for magnonic quantum information science (QIS) because of its extremely low damping. We report ultralow damping at 2 K in epitaxial $Y_3Fe_5O_{12}$ thin films grown on a diamagnetic $Y_3Sc_2Ga_3O_{12}$ substrate that contains no rare-earth elements. Using these ultralow damping YIG films, we demonstrate for the first time strong coupling between magnons in patterned YIG thin films and microwave photons in a superconducting Nb resonator. This result paves the road towards scalable hybrid quantum systems that integrate superconducting microwave resonators, YIG film magnon conduits, and superconducting qubits into on-chip QIS devices.

$Y_3Fe_5O_{12}$ (YIG) is a well-known ferrimagnetic insulator with extremely low magnetic damping, which makes it one of the best materials for fundamental studies and potential applications in magnonics, spintronics, and QIS.[1-8] To date, mm-scale YIG single-crystal spheres have been the material of choice for coherently coupling magnons to superconducting qubits[4] as well as for quantum sensing.[4, 7, 9, 10] For QIS applications that require scalable on-chip integration, ultralow-damping magnetic films, which can be patterned and integrated in hybrid quantum systems, are highly desired. However, there is a major barrier for using YIG thin films in QIS because epitaxial YIG films are mostly grown on $Gd_3Ga_5O_{12}$ (GGG) or other garnet substrates containing rare-earth elements, which induce excessive damping loss in YIG films across the interface at low temperatures.[11] If this substrate-induced damping enhancement can be eliminated while maintaining high YIG film quality, it will enable scalable on-chip QIS devices based on ultralow-damping YIG films operating at mK temperatures.

Previously, we have shown that a diamagnetic epitaxial buffer layer of $Y_3Sc_{2.5}Al_{2.5}O_{12}$ can effectively separate the magnetic coupling between the YIG film and GGG substrate, resulting in much improved damping at low temperatures.[12] Here, we demonstrate the growth of YIG epitaxial films on a new diamagnetic $Y_3Sc_2Ga_3O_{12}$ (YSGG) single-crystal substrate containing no rare-earth elements, which exhibit extremely low damping that decreases rapidly with temperature below 5 K. By integrating patterned YIG epitaxial films on YSGG with superconducting Nb coplanar waveguide (CPW) microwave resonators, we observe strong coupling between microwave photons and magnons at 2 K.

YIG thin films are epitaxially grown on YSGG (111) substrates using off-axis sputtering[13] at a substrate temperature of 675°C. The bulk lattice constant of YSGG is $a$ = 12.466 Å, which causes a 0.72% tensile strain in the YIG film (bulk $a$ = 12.376 Å). The crystalline quality of the YIG/YSGG films is characterized by X-ray diffraction (XRD) and X-ray reflectivity (XRR), as shown in Figs. 1a and 1b, respectively. The clear Laue oscillations of the YIG (444) peak indicate highly coherent crystalline ordering. The XRR peaks reveal the film thickness (30 nm) and low interfacial and surface roughness.

Broadband ferromagnetic resonance (FMR) measurement is performed on a YIG(30 nm)/YSGG(111) film at various temperatures ($T$) between 2 and 300 K in a Physical Property Measurement System (PPMS) from Quantum Design, as described previously.[12] We obtain FMR absorption spectra as a function of an in-plane magnetic field at various microwave frequencies ($f$) and temperatures. Figure 2a shows a representative derivative FMR spectrum at $T = 2$ K and $f = 4.3$ GHz. The resonance field $H_r$ and linewidth $\Delta H$ are extracted from fitting the FMR spectrum using $y = [A - B(H - H_r)]/[(H - H_r)^2 + (\Delta H/2)^2]^2$, where $A$ and $B$ are the symmetric and antisymmetric amplitudes of the lineshape, respectively. From Fig. 2a, we obtain $H_r = 710$ Oe and $\Delta H = 13.2$ Oe ($\sqrt{3} \times$ of peak-to-peak linewidth of 7.6 Oe).

We also measure the frequency-dependent FMR absorption at 2 K with a fixed in-plane field of 710 Oe using a vector network analyzer (VNA), as shown in Fig. 2b. The resonance frequency $f_r$ and linewidth $\Delta f$ are extracted from fitting the spectrum with a Lorentzian function $y = A/[(f - f_r)^2 + (\Delta f/2)^2]$. The obtained $f_r = 4.307$ GHz and $\Delta f = 36.3$ MHz agree well with the field linewidth $\Delta H = 13.2$ Oe as related by the gyromagnetic ratio of free electron, $\gamma/2\pi \approx 2.8$ MHz/G, where 36.3 MHz corresponds to 13.0 Oe.

Figure 2c shows the extracted field linewidths of the YIG(30 nm)/YSGG film as a function of frequency at $T = 2$ to 300 K. As temperature decreases from 300 K, the linewidth and the slope of their frequency dependence increase and peak at 20-30 K, after which both $\Delta H$ and the slope decrease quickly down to 2 K (limited by our PPMS). This behavior agrees with the slow-relaxation theory in YIG,[14-16] where the existence of rare-earth impurities in YIG induces a significant enhancement of relaxation at temperatures of 10's K. At temperatures below this regime, the linewidth and the slope decrease significantly.

According to the Landau-Lifshitz-Gilbert (LLG) equation,[17] the FMR linewidth is linearly dependent on the microwave frequency with the slope determined by the phenomenological Gilbert damping coefficient ($\alpha$): $\Delta H = 4\pi\alpha f/\gamma + \Delta H_0$, where $\Delta H_0$ is the inhomogeneous broadening which is generally attributed to magnetic nonuniformity[18] and surface defects.[19] We note that the LLG equation is a simplified theory and does not explain the nonlinearity observed in the frequency dependence of $\Delta H$, particularly at low temperatures, which requires additional contributions such as the slow-relaxation mechanism. Nonetheless, the phenomenological damping coefficient α from the LLG equation can serve as a figure of merit for a magnetic material.

We fit the linewidth vs. frequency data in Fig. 2c using the LLG equation and extract damping $\alpha$ and inhomogeneous broadening $\Delta H_0$ for each temperature, as shown in Fig. 2d. The damping starts at a very low value of $\alpha = 3.3 \times 10^{-4}$ at room temperature and increases as temperature drops. After reaching a peak of $\alpha = 1.3 \times 10^{-3}$ at 45 K, the damping decreases quickly at lower temperatures. Remarkably, the FMR linewidths are essentially frequency independent at 5 and 2 K, indicating an essentially zero damping according to the LLG equation. The rapid decrease of linewidth below 10 K implies that $\Delta H$ will likely decrease even further at sub-Kelvin temperatures.

The extremely low damping of the YIG epitaxial films on diamagnetic YSGG substrates at very low temperatures is highly promising for QIS studies. As an initial step in this regard, we integrate the YIG/YSGG films with superconducting resonators for the study of coupling between magnons in a YIG film and microwave photons emitted by a superconducting CPW resonator. First, we pattern a YIG(30 nm)/YSGG film into 10-μm wide strips using photolithography and Ar ion milling. Then, 300-nm thick Nb resonators are deposited by sputtering on the YSGG substrate with photolithography patterns followed by lift-off such that the YIG strips lie within the gaps of the CPW resonators.

The microwave resonator design includes a main bus channel which is capacitively coupled to two resonators, as shown in Fig. 3a. The two resonators are 13- and 13.5-mm long with both ends open-circuit, so they resonate with half-wavelength standing waves, as determined by $f_r = c/2l\sqrt{\epsilon_{\text{eff}}}$, where $c$ is the speed of light, $l$ is the length of the resonator, and $\epsilon_{\text{eff}}$ is the average dielectric constant of vacuum and YSGG substrate. The center conductor of the CPW resonators is 20-μm wide with a spacing gap of 15-μm from the group conductor on either side. A YIG strip is located within a gap of each resonator near the middle of the resonator length, as shown in the insets of Fig. 3a, which enables the magnetic field component of the resonating microwave to drive the FMR of the YIG strip. We fabricate two such devices with a total of four Nb resonators coupled to four YIG strips with 10-μm width and lengths of 300, 600, 900, and 1200 μm. In addition, we fabricate another such device with two Nb resonators on YSGG without YIG strips as a reference sample. The devices are mounted on a home-made sample holder and wire-bonded as shown in the insets of Fig. 3a, which is loaded into the PPMS and cooled down to 2 K.

We measure microwave transmission spectra of these devices using a VNA with an output power of -40 dBm, where each spectrum is averaged over 40 scans. Figure 3b shows the transmission spectrum (S21) of a device with two Nb resonators and no YIG, which exhibits two sharp dips at $f$ = 4.203 and 4.364 GHz, corresponding to the resonances of the 13.5- and 13-mm resonators, with high quality factors (Q) of 56,800 and 49,000, respectively. It is noted that the measured resonance frequency ratio of (4.364 GHz)/(4.203 GHz) = 1.0383 is almost identical to the length ratio of (13.5 mm)/(13 mm) = 1.0385, validating the resonator design and fabrication.

Figure 3c compares the transmission spectra of a bare Nb resonator (no YIG) and a Nb resonator coupled to a $10 \times 1200$ μm² YIG strip at zero field or in the presence of a magnetic field of 550 Oe, while Fig. 3d shows similar spectra for a Nb resonator coupled to a $10 \times 600$ μm² YIG strip. All Nb resonators with or without coupling to YIG strips exhibit high quality factors between 44,000 and 72,700 at zero magnetic field. To evaluate the performance of the Nb resonators in a magnetic field needed for strong microwave photon-magnon coupling, we apply an in-plane field of 550 Oe, which lowers Q by a factor of ~10×, although the quality factor remains quite high (>4,000). This behavior together with shift of resonance frequency due to the applied field and coupling to YIG are commonly seen in other microwave photon–magnon coupling reports.[20, 21]

Such resonator-ferromagnet hybrid systems can be modeled as a macrospin coupled to an *LC* resonator by the radio-frequency (rf) magnetic field component $b_{rf}$.[20] The eigenfrequencies of this system can be calculated as,

$$\omega_{\pm} = (\omega_r + \omega_m(H))/2 \pm \sqrt{(\omega_m(H) - \omega_r)^2 + 4g^2}/2, \qquad (1)$$

where $\omega_r$ is the resonance frequency of a standalone microwave resonator, $\omega_m(H)$ is the standalone ferromagnet's FMR frequency (dependent on the applied field $H$), and $g$ is the total photon-magnon coupling strength. The total coupling strength scales with the number of spins $N$ in the ferromagnet as $g = g_s\sqrt{N}$, where $g_s$ is the coupling strength between microwave photons and individual spins in the ferromagnet, which depends on the device design and microwave magnetic field strength $b_{rf}$ at the ferromagnet. Note that the number of spins here is a net value due to YIG's ferrimagnetism.

We perform field-dependent transmission measurement on the 13- and 13.5-mm Nb resonators to quantify their coupling strength to the $10 \times 300$ μm², $10 \times 600$ μm², $10 \times 900$ μm², and $10 \times 1200$ μm² YIG strips. Figures 4a-4d show the microwave transmission of the four resonator-YIG hybrid devices as a function of field and frequency measured at 2 K. The anticrossing features in the plots are the signatures of microwave photon-magnon coupling, where YIG's FMR frequency $\omega_m(H)$ meets the microwave resonator's resonance frequency $\omega_r$ and the degeneracy is lifted. By combining Eq. (1) and the Kittel equation, $\omega_m(H) =$ $\gamma\sqrt{H(H + 4\pi M_{\text{eff}})}$, where $M_{\text{eff}}$ is the effective saturation magnetization of the YIG film, we obtain the angular frequencies of the hybrid systems' modes,

$$\omega_\pm = (\omega_r + \gamma\sqrt{H(H + 4\pi M_{\text{eff}})})/2 \pm \sqrt{\left(\gamma\sqrt{H(H + 4\pi M_{\text{eff}})} - \omega_r\right)^2 + 4g^2}/2 \qquad (2)$$

We point out that different aspect ratios of the YIG strips can give rise to different demagnetizing factors, which slightly shift $\omega_m(H)$. However, this shift is smaller than 1 Oe over the field and frequency range of interest and thus is ignored here. We fit the anticrossing features with Eq. (2) and extract the coupling strength $g$ which is dependent on the magnetic volume and the number of spins $N$. For the fitting, a background linearly dependent on the field is added to Eq. (2) to account for the field-induced microwave resonator frequency shift.

To determine the number of spins in a YIG strip, we measure magnetic hysteresis loops for the YIG(30 nm)/YSGG film using a superconducting quantum interference device (SQUID) magnetometer at 2 and 300 K, as shown in Fig. 4e. The saturation magnetization is determined to be $M_S = 180$ emu/cm$^3$ at 2 K and 131 emu/cm$^3$ at 300 K. The hysteresis loops remain largely square with a small coercivity of 1.1 and 3.1 Oe at 300 and 2 K, respectively, corroborating the high magnetic uniformity at both room and cryogenic temperatures.

Figure 4f shows the linear dependence of coupling strength $g/2\pi$ on the square root of both the YIG volume and the number of spins. From Fig. 2b, the decay rate $\kappa_m/2\pi$ of YIG's resonance mode near 4.3 GHz at 2 K and $H = 710$ Oe is determined to be 36.3 MHz. Next, the decay rate $\kappa_r/2\pi$ of the Nb resonator is estimated to be 1.037 MHz based on the transmission linewidth of the resonator coupled to the 10 × 600 μm$^2$ YIG strip at $H = 550$ Oe as shown in Fig. 3d. Thus, the cooperativity for the 10 × 600 μm$^2$ YIG is calculated to be $g^2/(\kappa_r\kappa_m) \approx$ 57.9, showing that the coupling strength is much stronger than the decay rates of individual components, which is required for coherent coupling. Using $M_S = 180$ emu/cm$^3$ at 2 K for the YIG/YSGG film, we determine the average coupling strength to an individual spin to be $g_s/2\pi$

= 25.1 Hz. This coupling strength $g_s/2\pi$ is comparable to other reports on microwave photon-magnon coupling using ferromagnetic metals[20-22] and similar CPW resonator-ferromagnet structures. Considering that the microwave magnetic field in the gap of a CPW (where our YIG strip is located) is weaker than that for a ferromagnet directly on top (or underneath) the superconducting center conductor channel, one can achieve stronger microwave photon-magnon coupling by positioning the YIG strip near a stronger microwave magnetic field or utilizing resonator designs such as lumped-element $LC$ resonator, which can provide far stronger coupling strength.[20]

This work provides the first demonstration that ultralow-damping YIG epitaxial films on YSGG can be integrated with superconductor resonators to achieve strong microwave photon-magnon coupling at few Kelvin temperatures. Such ultralow-damping YIG films offer clear advantages (in terms of decay rate) over metallic ferromagnets for on-chip hybrid quantum systems that incorporate magnonic conduits, microwave superconductor resonators, and superconductor qubits for QIS applications that operate in the mK regime.

In summary, we demonstrate the growth of high-quality epitaxial YIG thin films on diamagnetic YSGG substrate, which exhibit narrow FMR linewidth and extremely low damping at 2 K. We couple these YIG films to superconducting Nb resonators to create hybrid structures that achieve strong microwave photon-magnon coupling. This work demonstrates the potential power of ultralow-damping YIG films for scalable, integrated QIS applications at low temperatures.

This work was primarily supported by the Center for Emergent Materials: an NSF MRSEC under award number DMR-2011876. D.R. acknowledges partial support from the National Science Foundation under award number DMR-2225646 (YIG film growth and X-ray characterizations).

**Figure Captions:**

**Figure 1**. **X-ray diffraction results of YIG films.** (a) $2\theta/\omega$ XRD scan and (b) XRR scan of a YIG(30 nm) epitaxial film grown on YSGG (111) substrate, indicating the high crystalline quality of the YIG film.

**Figure 2**. **FMR measurements of a YIG(30 nm)/YSGG film.** (a) Field-dependent derivative FMR spectrum at 2 K driven by a microwave frequency of 4.3 GHz. (b) Frequency-dependent FMR spectrum at 2 K with an in-plane field of 710 Oe. (c) Frequency dependence of FMR linewidth at different temperatures, from which the damping constant and inhomogeneous broadening are obtained by linear fitting. (d) Temperature dependence of damping (red) and inhomogeneous broadening $\Delta H_0$ (blue). The error bars are from the linear fitting in (c).

**Figure 3**. **Microwave transmission of Nb CPW resonators with or without YIG strips and magnetic field at 2 K.** (a) Schematic of Nb resonator device with YIG strips placed within their gaps. The size of the whole device is $3.5 \times 4.4$ mm$^2$. The lengths of the two Nb resonators are 13 mm and 13.5 mm. Insets: optical microscope images of selected areas with the same magnification. The YIG strips shown (color contrast augmented) are $10 \times 900$ μm$^2$ (top) and $10 \times 300$ μm$^2$ (bottom). (b) Microwave transmission (S21) spectrum of two Nb resonators without a YIG strip in their gaps. The two sharp dips (resonances) at 4.364 and 4.203 GHz correspond to the resonance frequencies of the 13 mm and 13.5 mm resonators, respectively. (c) Microwave transmission spectra of the 13.5 mm resonators without YIG at zero field (blue), with a $10 \times 1200$ μm$^2$ YIG strip at zero field (orange), and with a $10 \times 1200$ μm$^2$ YIG strip in the presence of a 550 Oe field (green). (d) Microwave transmission spectra of the 13 mm resonators without YIG at zero field (blue), with a $10 \times 600$ μm$^2$ YIG strip at zero field (orange), and with a $10 \times 600$ μm$^2$ YIG strip at 550 Oe (green).

**Figure 4. Microwave photon-magnon coupling in Nb resonator-YIG(30 nm) hybrid structures on YSGG.** Microwave transmission in Nb resonators coupled to YIG strips of (a) 10 × 300 μm$^2$, (b) 10 × 600 μm$^2$, (c) 10 × 900 μm$^2$, and (d) 10 × 1200 μm$^2$, as a function of microwave frequency and in-plane magnetic field at 2 K. (e) Magnetic hysteresis loops of a YIG(30 nm)/YSGG film at 2 and 300 K. (f) Coupling strength between microwave photons and magnons in the YIG films as a function of the square root of the magnetic volume as well as the number of spins in YIG.

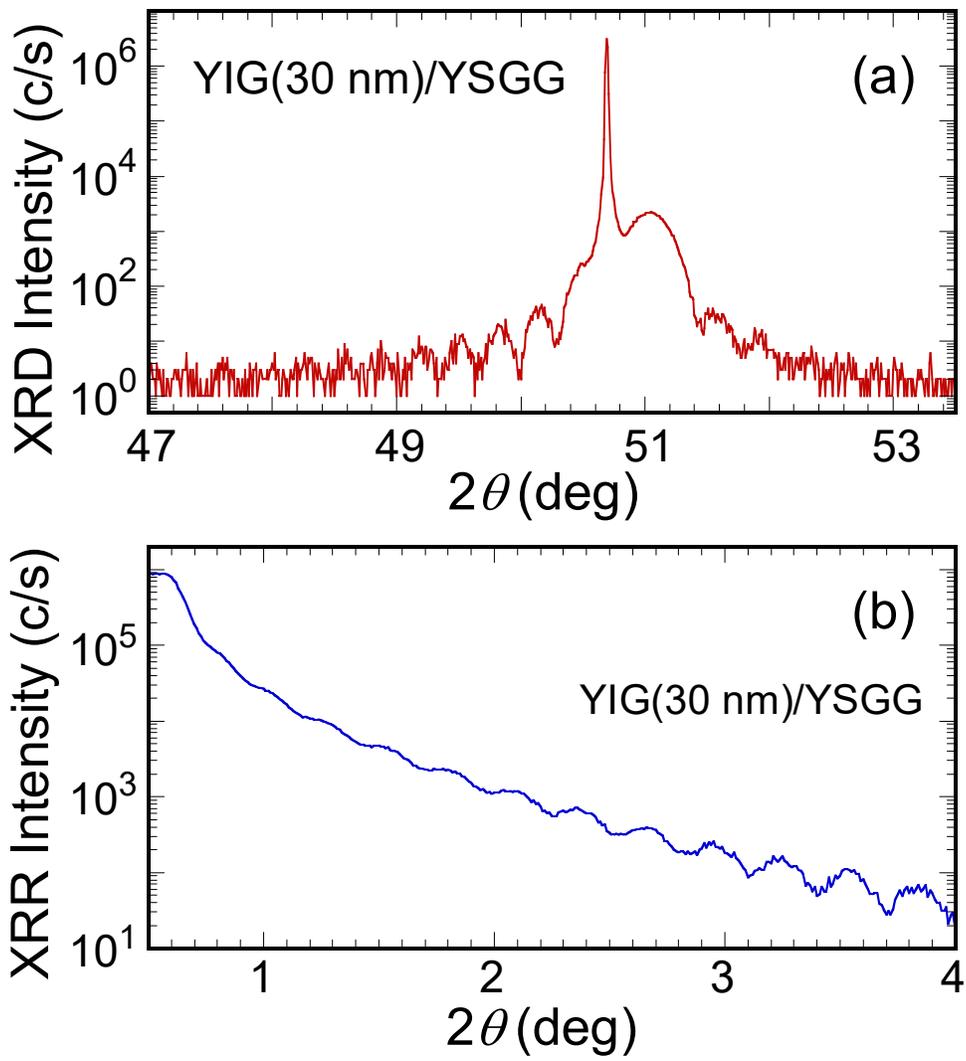

**Figure 1.**

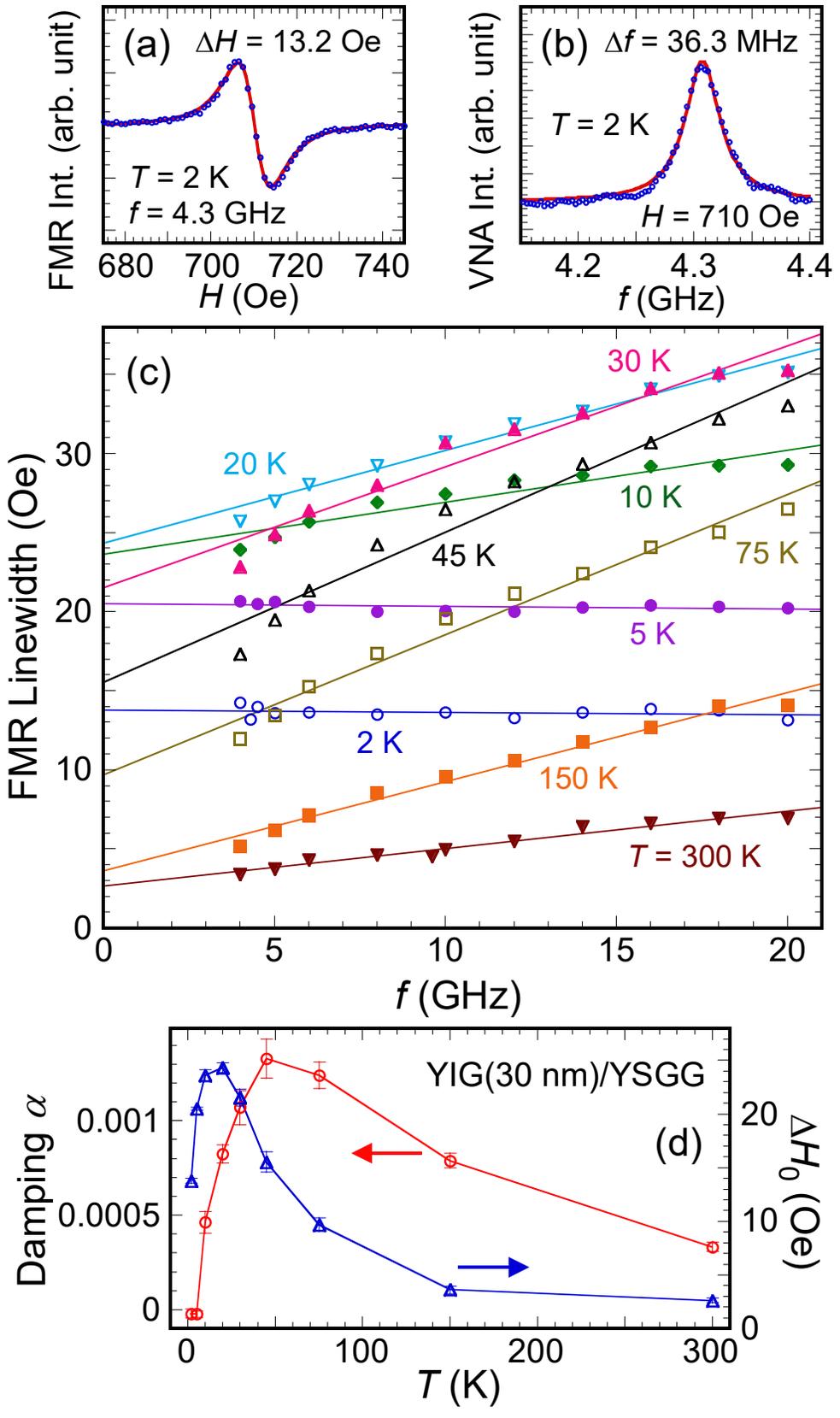

**Figure 2.**

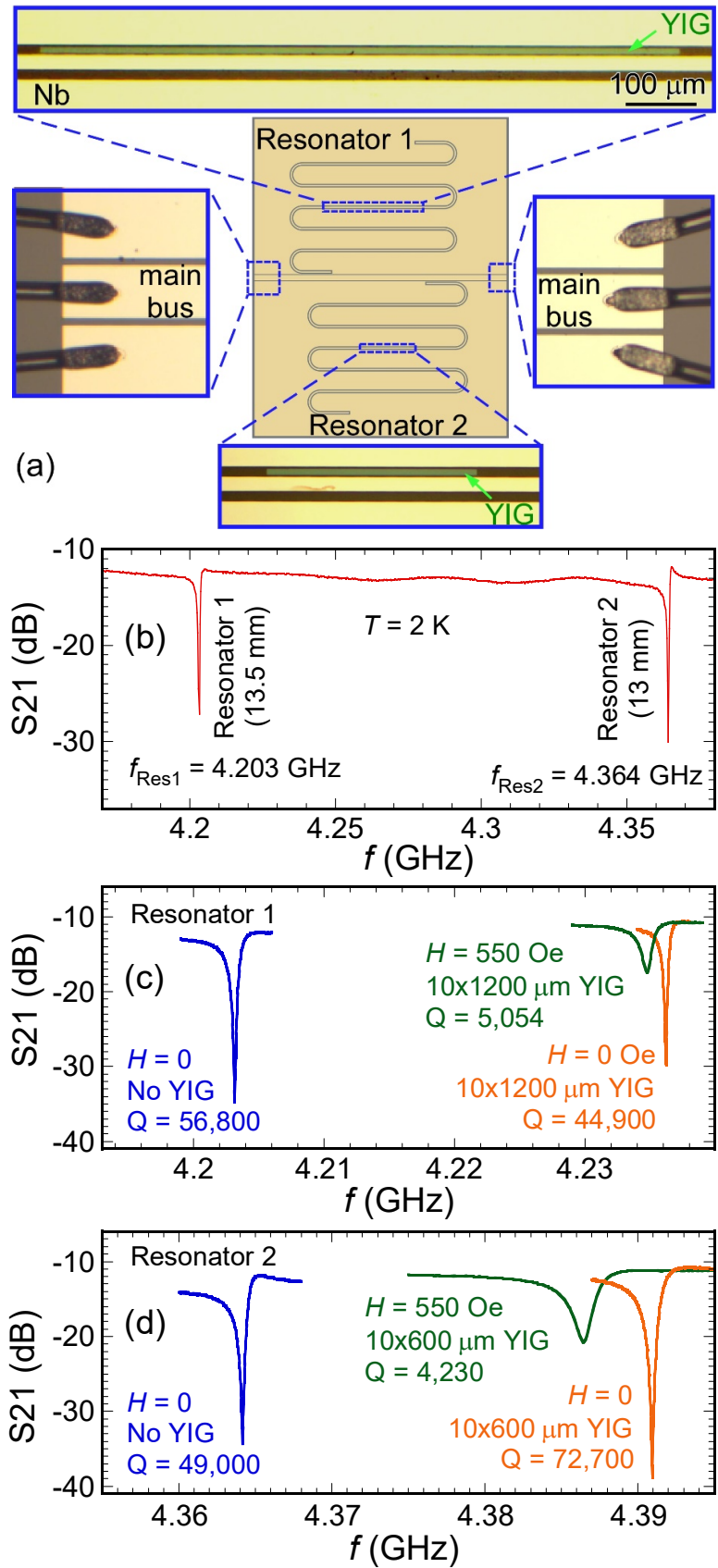

**Figure 3.**

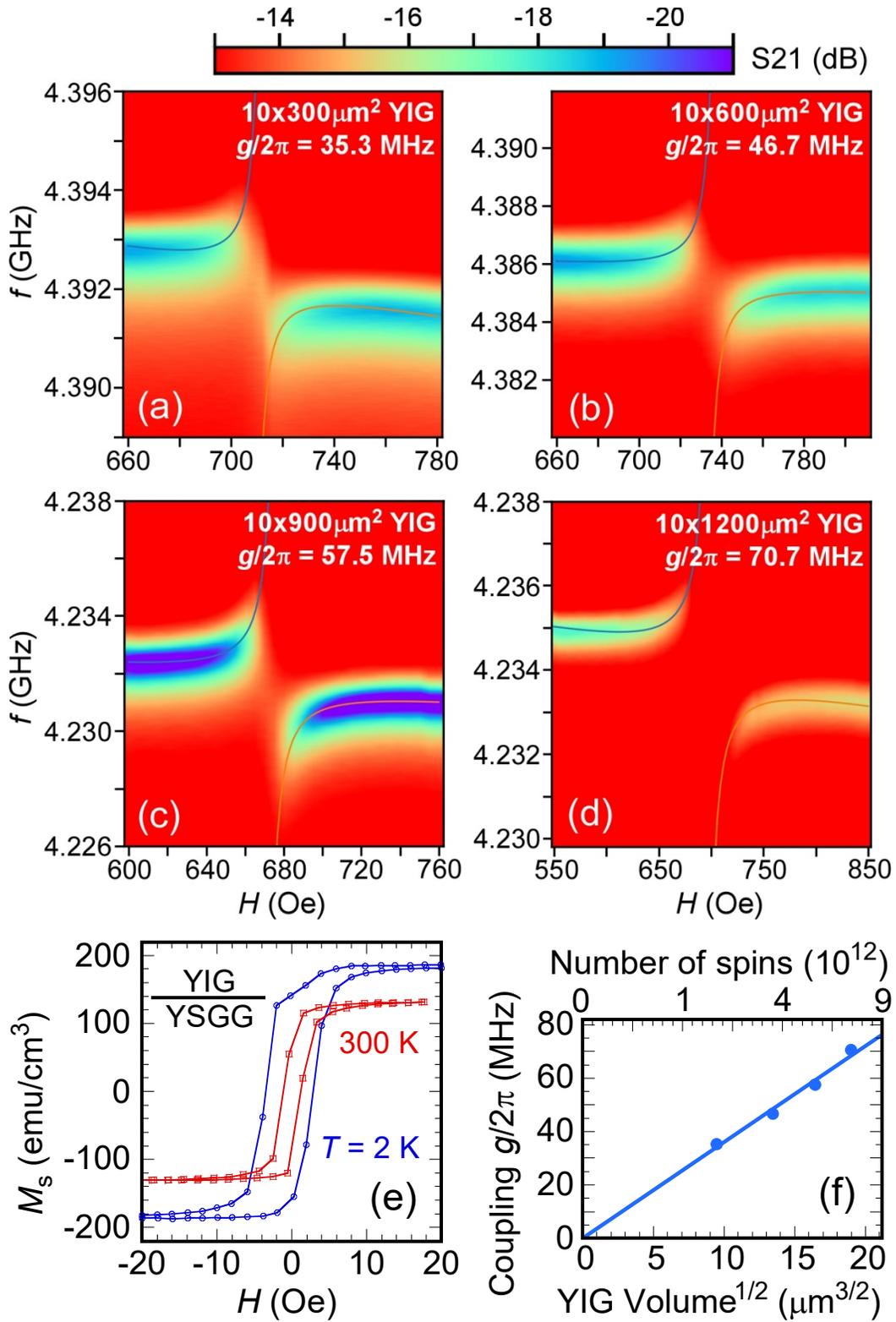

**Figure 4.**